\documentclass[preprint,authoryear,12pt]{elsarticle}

\usepackage{color}
\newcommand{\red}[1]{\textcolor{black}{#1}}
\newcommand{\blue}[1]{\textcolor{black}{#1}}
\newcommand{\green}[1]{\textcolor{black}{#1}}
\usepackage{dcolumn}
\newcolumntype{d}[1]{D{.}{\cdot}{#1}}
\newcolumntype{.}{D{.}{.}{-1}}
\newcolumntype{,}{D{,}{,}{-1}}

\usepackage{graphicx}

\usepackage{amssymb,multirow}

\journal{Journal of Theoretical Biology}

\begin{document}

\begin{frontmatter}


\title{
Transcription fluctuation effects on biochemical oscillations
}

\author{Ryota Nishino, Takahiro Sakaue, and Hiizu Nakanishi}

\address{
Department of Physics, Kyushu University 33, Fukuoka 812-8581, Japan
}

\begin{abstract}
Some biochemical systems show oscillatory behavior.  These systems often
consist of negative feedback loops with repressive transcription
regulation.  Such biochemical systems have distinctive characteristics
in comparison with ordinary oscillatory chemical systems:
i) the numbers of molecules involved are small, 
ii) there are typically only a couple of genes in a cell with a finite
regulation time scale.
Due to the fluctuations caused by these features, the system behavior
can be quite different from the one obtained by deterministic rate
equations, because the rate equations ignore molecular fluctuations and
thus are exact only in the infinite molecular number limit.
The molecular fluctuations on a free-running circadian system have been
studied by Gonze et al. (2002) 
\red{by introducing a scale parameter $\Omega$ for the system size.  }
They consider, however, only the first effect, assuming that the gene
process is fast enough \blue{ for the second effect to be ignored, but
this has not been examined systematically yet.}
In this work, we study fluctuation effects due to the finite 
gene regulation time
\red{ by introducing a new scale parameter $\tau$, which we take as the
unbinding time of a nuclear protein from the gene.  We particularly
focus on}
the case where the fluctuations due to small molecular numbers can be
ignored.  In simulations on the same system \red{studied} by Gonze et
al., \red{ we find the system is unexpectedly sensitive to the
fluctuation in the transcription regulation; the period of oscillation
fluctuates about 30 min even when the regulation time scale $\tau$ is
around 30 s,} 
that is even smaller than 1/1000 of its circadian period.
We also demonstrate
\red{that the distribution width for the oscillation period and the
amplitude scales with $\sqrt\tau$, and the correlation time of the
oscillation scales with $1/\tau$ in the small $\tau$ regime.}
The relative fluctuations for the period are about half of that for the
amplitude, namely, the periodicity is more stable than the
amplitude.
%

\end{abstract}


\end{frontmatter}



\section*{Introduction}

One of the outstanding features in biological systems is that the
systems often operate on surprisingly small numbers of active
molecules, yet they seem to work quite reliably.
This is especially intriguing in the case where the chemical reaction
system involves a gene transcription 
because there are typically only a couple of genes in a cell.

One example is a circadian system, which shows a rhythmic
behavior of approximately 24-hour periodicity.  It is a universal
feature of biological systems and known to be very accurate and robust
against external and internal
perturbations \citep{Dunlap-1999,Young-2000}. Its biochemical mechanisms
have been proposed in several
systems \citep{Goldbeter-1995,Goldbeter-1996,Leloup-1999}, and most of
them are based on a time-delayed negative feedback loop of a biochemical
reaction network which includes transcription regulations.  Some of the
protein molecules are expected to be very small in number, and the
%
number of each gene is typically of the order of one in a cell and does
not scale with the cell size, 
thus it is surprising that the circadian system is capable of
maintaining its extraordinary regularity especially in the case of a
single cell organism \citep{Barkai-2000}.

The effects of molecular fluctuations on the circadian system has been
studied by  
\cite{Gonze-2002-a,Gonze-2002-b} by Monte Carlo simulations using the
Gillespie method \citep{Gillespie-1976,Gillespie-1977} with the scale
parameter $\Omega$ for the molecular numbers.  By simulating the system
with various values of the scaling parameter $\Omega$, they demonstrated
that the system shows reasonably coherent oscillation as long as the
system contains more than several tens of mRNA, thus concluded that
their system are fairly robust against molecular fluctuations.

They examined the system rather systematically based upon a standard
method to study the stochastic nature of chemical
reactions \citep{Nicolis-1972} by scaling the reaction rates in the way
to keep the rate equation unchanged.  However, there is an ambiguity in
the treatment of the gene regulation process because the number of
genes should not scale with other protein numbers.  
They scaled the reaction rates
\red{proportional to $\Omega$ for the gene processes, namely,
they made the gene regulation times infinitely fast
}
in the large $\Omega$ limit, \red{thus} the system dynamics reduce to
the one described by the corresponding rate equations \red{without the
gene processes} \citep{Gonze-2002-a}.
\blue{This could be justified only when the gene processes are so fast
that they do not cause significant fluctuation on the system behavior.
}
%
In fact, it is not \red{reasonable} to assume that the time scale of the
gene regulation depends upon the scale parameter $\Omega$, if you think
of it as the cell volume, because the time scale with which a regulatory
protein binds to the operator site is determined by the protein
concentration, and the time scale with which the protein unbinds is
determined by its binding energy.

In this work, \blue{in order to analyze fluctuations from the
two distinct origins separately,  we introduce a new
scale parameter $\tau$ in addition to $\Omega$.
The scale paramter $\tau$} scales the binding/unbinding time of the
gene regulatory protein. 
%
%
Thus, these two parameters, $\tau$ and $\Omega$, control the two
distinct fluctuation sources that exist in the biological systems,
namely, $\Omega$ controls the fluctuations \red{due to the finite}
molecular numbers while $\tau$ controls the fluctuations \red{due to
the finite gene regulation times.}
%
%
We perform the Monte Carlo simulations on the same system
as the one studied by
\cite{Gonze-2002-b}, focusing on the latter effect,
\blue{ and demonstrate that significant fluctuation can arise from the
stochasticity in the gene process alone.  We also examine how the
fluctuation scales with  $\tau$. }


\section*{Model}

\begin{figure}
\centerline{\includegraphics[width=8cm]{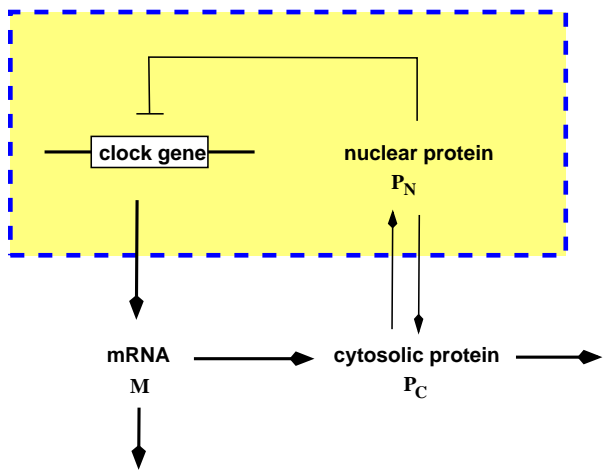}}
\caption{
Simplified core model for a circadian system.}
\label{Circadian}
\end{figure}

The model we study is the simplest version of core model for a circadian
system that consists of a gene G, mRNA M, cytosolic protein $\rm P_C$,
and nuclear protein $\rm P_N$ (Fig.\ref{Circadian}).  The biochemical
reactions for these elements are given by
\begin{eqnarray}\rm
G + {\it n} P_N & \rightleftarrows &\rm GP_{N\it n},
\\\rm
G & \rightarrow &\rm G + M,
\\\rm
M & \rightarrow &\rm \times,
\\\rm
M & \rightarrow &\rm M + P_C,
\\\rm
P_C & \rightarrow &\rm \times,
\\\rm
P_C & \rightleftarrows &\rm P_N.
\end{eqnarray}
The parameter $n$ is the number of nuclear proteins $\rm P_N$ that bind
to suppress the gene, i.e. Hill coefficient for the gene activity; we
adopt $n=4$ for most of the calculation 
\blue{as \cite{Gonze-2002-a,Gonze-2002-b}.}

Now, we introduce \red{the two scaling parameter $\Omega$ and $\tau$;
$\Omega$ scales the reaction rates so that the numbers of mRNA and the
proteins become proportional to it, and $\tau$ scales the
binding/unbinding time of the nuclear protein to the gene operator site.}
The transition rates for each reaction are listed in
Table \ref{table-I}, where we define the variables $G$, $M$, $P_C$, and
$P_N$ to represent the numbers of active genes, mRNA, cytosolic
proteins, and the nuclear proteins in a single cell, respectively.  The
gene variable $G$ takes either 1 or 0 \blue{values}, depending upon the
active state (G) or the inactive state ($\rm G P_{N\it n}$),
respectively.
Note that we employ Michaelis-Menten enzymatic reactions for 
the degradation processes.
\red{
The first two reaction rates in Table \ref{table-I} are for the gene
regulation and proportional to $1/\tau$, but do not scale with $\Omega$}
because we assume only one gene in a cell\footnote{
%
\cite{Gonze-2002-a} used $\Omega$ as a scaling parameter
in the places where we use $1/\tau$.
}.  
\blue{
The ratio of the binding and the unbinding
times is determined in the way that the corresponding average
behavior described by the rate equation remains the same with the
original system 
in the small $\tau$ limit.}
%
%
On the other hand, the gene transcription activity in the third reaction
is scaled as $v_s\Omega$ in order that the numbers of mRNA and the
proteins should be proportional to $\Omega$.

If we naively write down differential equations for the time evolution,
ignoring the fact that the variables are integers, we would have
\begin{eqnarray}
\tau{dG\over dt} &  = & (1-G) - \left({P_N/\Omega\over K_I}\right)^n G,
\\
{dM\over dt} & = & v_s\Omega\, G - v_m\Omega{M\over K_m\Omega+M},
\\
{dP_C\over dt} & = & 
k_s M - v_d\Omega {P_C\over K_d\Omega+P_C} -k_1 P_C + k_2 P_N,
\\
{dP_N\over dt} &  = & k_1 P_C - k_2 P_N
\\&&\qquad
+{n\over\tau}\left[(1-G) - \left({P_N/\Omega\over K_I}\right)^nG\right].
\label{P_N-eq}
\end{eqnarray}
For ordinary chemical reactions without a gene transcription process,
the stochastic dynamics should be well described by such equations
for the large $\Omega$ case, where the numbers of molecules are large.
However, in the present system, the number of gene is one and does not
scale with $\Omega$, thus the stochastic nature remains even in the case
of the infinite $\Omega$ as long as $\tau$ is finite.

\subsection*{Large $\Omega$ limit:}

Supposing the scale parameter $\Omega$ as a cell volume,
we define the ``concentrations'' of mRNA and the proteins as,
\begin{equation}
{\rm [M]}\equiv {M\over\Omega},\quad
{\rm [P_C]}\equiv {P_C\over\Omega},\quad
{\rm [P_N]}\equiv {P_N\over\Omega},
\end{equation} 
and write down the rate equations for them as
\begin{eqnarray}
{d{\rm [M]}\over dt} & = & 
v_s\, G - v_m{{\rm [M]}\over K_m+{\rm [M]}},
\label{rate_M}
\\
{d{\rm [P_C]}\over dt} & = & 
k_s {\rm [M]} - v_d {{\rm [P_C]}\over K_d+{\rm [P_C]}}
 -k_1 {\rm [P_C]} + k_2{\rm [P_N]},
\label{rate_P_C}
\\
{d{\rm [P_N]}\over dt} &  = & k_1 {\rm [P_C]} - k_2 {\rm [P_N]},
\label{rate_P_N}
\end{eqnarray}
where we have ignored the term of the order of $n/\Omega$
in Eq.(\ref{rate_P_N}).

For ordinary chemical reactions, we expect that the deterministic
dynamics represented by the rate equations would describe the
system accurately in the large $\Omega$ limit, because the effect of
molecular fluctuation becomes negligible.  However, for the present
case, the system remains stochastic even in the large $\Omega$ limit
because 
the variable $G$ remains stochastic.

\subsection*{Small $\tau$ limit:}

In the case where $\tau$ is much smaller than any other time scales in
the system, \red{the system reduces to the one studied by
\cite{Gonze-2002-b}.
This can be seen by
introducing} the time dependent average value of $G$,
denoted by $G_{\rm Av}(t)$, i.e. the time average of $G$ over the
longer time scale than $\tau$ but shorter than other time scales in the
system.  Its value is given by the condition that the first two
reactions in Table \ref{table-I} are equilibrated,
\begin{eqnarray}
G_{\rm Av}(t) & = & {1\over 1 + ({\rm [P_N]}/K_I)^n}.
\label{G_Av}
\end{eqnarray}
Then the system dynamics are given by the stochastic dynamics of
\red{reaction $1\sim 6$} in Table
\ref{table-I} with $G$ replaced by $G_{\rm Av}$.

If we take the large $\Omega$ limit on top of this, we obtain the
deterministic rate equations given by
Eqs.(\ref{rate_M})$\sim$(\ref{rate_P_N}) with $G$ being replaced by
$G_{\rm Av}(t)$ of Eq.(\ref{G_Av}).

\begin{table}[h]
\centerline{\begin{tabular}{c@{\hspace{.5cm}}c c}
\hline
no. & reaction &  transition rate
\\\hline\\
a & \parbox{1.1cm}{$G=1$\\ $P_N$}  $\longrightarrow$  
\parbox{1.5cm}{$G=0$\\$P_N -n$} &
$\displaystyle {1\over\tau}\,\left({P_N\over K_I\Omega}\right)^n G$ 
\\\\
b & \parbox{1.1cm}{$G=0$\\ $P_N$}  $\longrightarrow$  
\parbox{1.5cm}{$G=1$\\$P_N +n$} & 
$\displaystyle {1\over\tau}\,(1-G)$
\\\\\hline\\
1 & $M$  $\longrightarrow$  $M+1$ & $v_s\Omega\, G$
\\\\
2 & $M$  $\longrightarrow$  $M-1$ &
$\displaystyle v_m\Omega\,{M/\Omega\over K_m+M/\Omega}$ 
\\\\
3 & $P_C$  $\longrightarrow$  $P_C+1$ &  $k_s M$ 
\\\\
4 & $P_C$  $\longrightarrow$  $P_C-1$ &
$\displaystyle v_d\Omega\,{P_C/\Omega\over K_d+P_C/\Omega}$ 
\\\\
5 & \parbox{0.5cm}{$P_C$\\$P_N$}  $\longrightarrow$  
\parbox{1.1cm}{$P_C-1$\\$P_N+1$} & 
$\displaystyle k_1 P_C$ 
\\\\
6 & \parbox{0.5cm}{$P_C$\\$P_N$}  $\longrightarrow$  
\parbox{1.1cm}{$P_C+1$\\$P_N-1$} & $k_2 P_N$ 
\\\\\hline
\end{tabular}}
\caption
{Reaction table for a simplified circadian system.}
\label{table-I}
\end{table}

\section*{Simulations and results}

In order to examine the effect of gene fluctuations, we have performed
numerical simulations for various values of $\tau$ and $\Omega$.  
We examine two cases: (i) the case where both $\tau$ and $\Omega$
are finite, and (ii) the case where $\tau$ is finite but in the large
$\Omega$ limit.
In the first case, the fully stochastic dynamics are given by Table
\ref{table-I}; for these we employ the Gillespie method
\citep{Gillespie-1976, Gillespie-1977}.  In the second case, the
concentrations $\rm [M]$, $\rm [P_C]$, and $\rm [P_N]$ follow the
deterministic dynamics while the gene process remains stochastic.  In
this case, we integrate the rate equations
(\ref{rate_M})$\sim$(\ref{rate_P_N}) using Runge-Kutta method, but at
every time step of the length $\Delta t$, the gene variable $G$ is
subject to a trial for change according to the probability $w\Delta t$
under Poisson process with $w$ being the transition rate given in the
first two processes in Table 1.

\begin{figure}
\centerline{\includegraphics[width=7cm]{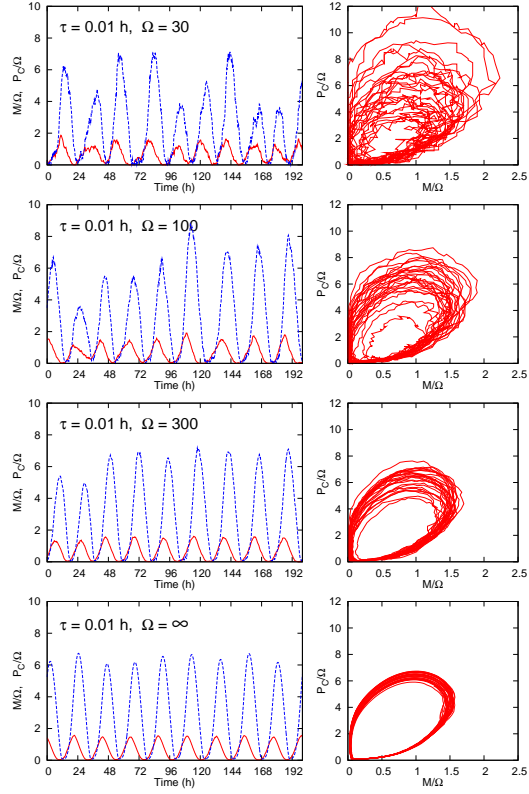}}
\caption{
Oscillatory behaviors of the concentrations of mRNA and the cytosolic
protein for various values of $\Omega$ with $\tau=0.01$ h.  The plots in
the left column show the time variation and those in the right column
are the projections of the trajectories in the 
$\rm M/\Omega - P_C/\Omega$ plane.
We employ the same reaction parameters with those in Gonze et al.:
$n=4$, $v_s=0.5$ nM h$^{-1}$, $K_I=2.0$ nM, $v_m=0.3$ nM h$^{-1}$,
$K_m=0.2$ nM, $k_s=2.0$ h$^{-1}$, $v_d=1.5$ nM h$^{-1}$,
$K_d=0.1$ nM, $k_1=0.2$ h$^{-1}$, $k_2=0.2$ h$^{-1}$.
}
\label{tau0.01-Omega}
\end{figure}

Figure \ref{tau0.01-Omega} shows the system behaviors for various values
of $\Omega$ with $\tau=0.01$h.  The other reaction parameters are the
same as those used by 
\cite{Gonze-2002-b}.  The plots in
the left column are the time variations of the concentrations of mRNA
(
solid lines) and the cytosolic protein (
dashed lines), and
those in the right column show the oscillation trajectories projected
on the $M/\Omega-P_C/\Omega$ plane of the phase space.  The fluctuation
decreases as $\Omega$ increases, but it remains finite even in the
infinite $\Omega$ case because of the fluctuation from the gene
activity.

\begin{figure}
\centerline{\includegraphics[width=7cm]{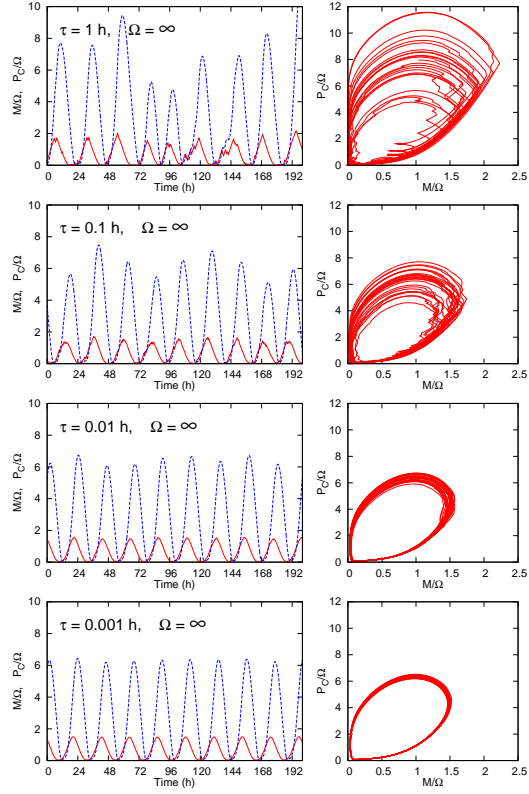}}
\caption{
Oscillatory behaviors of the concentrations of mRNA and the cytosolic
protein for various values of $\tau$ with $\Omega=\infty$.
The parameters are the same with those in Fig.\ref{tau0.01-Omega}.
}
\label{OmegaInfy}
\end{figure}

In order to see the effect of gene stochasticity, we examine the case
for various values of $\tau$ in the large $\Omega$ limit
(Fig.\ref{OmegaInfy}).  The fluctuation decreases on decreasing $\tau$
as in the case of increasing $\Omega$.  The trajectories are smoother
in comparison with the previous case because the stochasticity is
limited to the gene activity.  One can see that the fluctuation is
evident even in the case $\tau=0.01$h, where the ratio $\tau$ to the
period ($\sim$ 24h) is as small as 0.5$\times 10^{-3}$.

\begin{figure}
\centerline{\includegraphics[width=6cm]{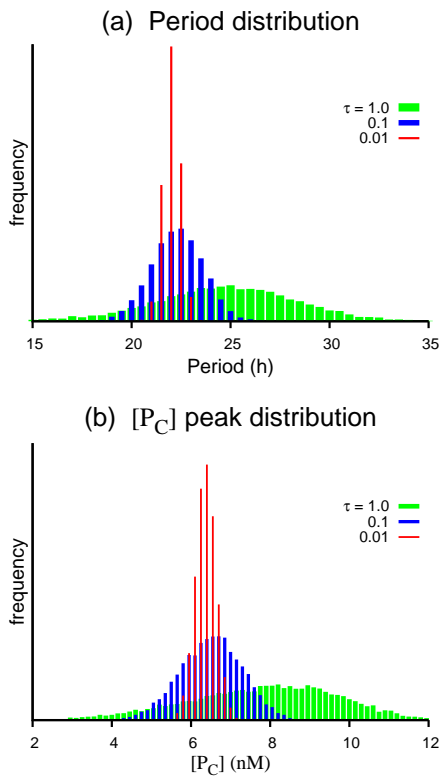}}
\caption{
Distributions for (a) the period (i.e. peak-to-peak interval) and (b)
the peak value of the cytosolic protein variation for $\tau=1.0$, 0.1,
0.01 h with $\Omega=\infty$.  The averages and the standard deviations
are tabulated in Table \ref{Table-II}, from which one can see that the
 width of the distribution scales roughly as $\sqrt\tau$.  } 
\label{dist}
\end{figure}

Figure \ref{dist} shows the period (i.e. the peak-to-peak interval)
distributions and the peak value distributions of $\rm [P_C]$ for
$\tau=1.0$, 0.1, and 0.01 h. The averages and the standard
deviations for the distributions are tabulated in Table \ref{Table-II}.
Both of the distributions becomes narrower for the smaller value of
$\tau$ approximately as $\sqrt\tau$, but the standard deviation of the
period distribution is still about a half hour even for the case of
$\tau=0.01$ h.  It should be noted that the ratios of the standard
deviation to the average for the peak value distributions are about
twice as large as those for the period distributions.
\red{More systematic data are presented in Supplementary material to
show the $\sqrt\tau$ scaling and the ratio of the two distribution widths.}


\begin{table}
\centerline{
\begin{tabular}{cc...}
\hline
$\tau$ (h) & & \hspace{4ex} 1\hspace{4ex}  &
 \hspace{3ex}0.1 \hspace{3ex}  & \hspace{2.5ex}0.01\hspace{2.5ex}  \\
\hline\hline
\multirow{3}{*}{period (h)}&
av. &  24.6   & 22.3  & 22.0  \\
\cline{2-5}
& std. &  3.92  & 1.32  & 0.45 \\
\cline{2-5}
& std./av. &  0.16  & 0.059  & 0.020 \\
\hline
\multirow{3}{*}{peak value (nM)}&
av. & 7.82   & 6.50   & 6.37  \\
\cline{2-5}
& std. &  1.87  & 0.79  & 0.26\\
\cline{2-5}
& std./av. & 0.24 & 0.12  & 0.041 \\
\hline
\end{tabular}}
\caption{
The averages and the standard deviations for the period and the peak
value distributions for the cytosolic protein concentration shown in
Fig.\ref{dist}.  
{The ratios of the standard deviation to the average
for the peak value distributions are about twice as large as those for
the period distributions.}
}
\label{Table-II}
\end{table}

\begin{figure}
\centerline{\includegraphics[width=7cm]{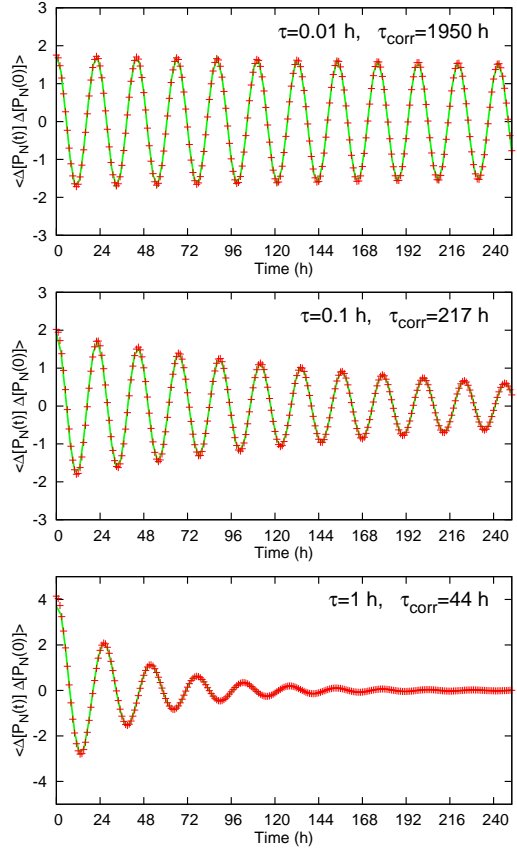}}
\caption{
The time correlation functions for the nuclear protein concentration for
the regulation time $\tau=$0.01, 0.1, 1 h with $\Omega=\infty$.
The (green) lines shows the fitting curves of the form
$A\cos(\omega_0 t+\theta_0)\,e^{-t/\tau_{\rm corr}}$.
The fitted values of $\tau_{\rm corr}$ are shown on the plots.
The other parameters are the same with those in Fig.\ref{tau0.01-Omega}.
} \label{corr}
\end{figure}

\begin{figure}
\centerline{\includegraphics[width=7cm]{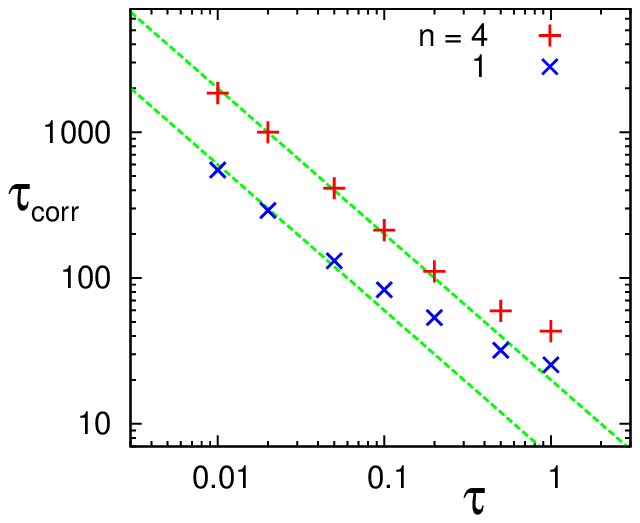}}
\caption{
The regulation time $\tau$ dependences of the correlation time
$\tau_{\rm corr}$ for $n=1$ and 4 with $\Omega=\infty$ in the
logarithmic scale.  The (green) dashed lines shows the fitted lines
proportional to $\tau^{-1}$.
}
\label{tau-tau_corr}
\end{figure}

The time correlation function $C(t)$ of the nuclear protein
concentration $\rm [P_N(t)]$ is defined as
\begin{equation}
C(t) = {1\over T}\int_0^T 
     \Delta {\rm [P_N(t'+t)]}\,\Delta {\rm [P_N(t')]}\, dt',
\end{equation} 
where $\Delta {\rm [P_N(t)]}$ represents the deviation from the average,
\begin{equation}
\Delta {\rm [P_N(t)]}\equiv {\rm [P_N(t)]}-
               {1\over T}\int_0^T  {\rm [P_N(t')]}\, dt'
\end{equation}
with $T$ being the time length of the whole simulation.
In Fig.\ref{corr}, the correlation functions are plotted and fitted to
 the form of damped oscillation
\begin{equation}
  A \cos(\omega_0 t +\theta_0)\, e^{-t/\tau_{\rm corr}}
\label{damped-osci}
\end{equation} 
to estimate the correlation time $\tau_{\rm corr}$.  Figure
\ref{tau-tau_corr} shows the $\tau$ dependence of the correlation time
$\tau_{\rm corr}$ in the logarithmic scale.  It shows the scaling
\begin{equation}
\tau_{\rm corr} \sim {1\over \tau}
\label{tau_corr}
\end{equation}
in the small $\tau$ regime, and the longer correlation time in the $n=4$
case than in the $n=1$ case.  One may notice that the correlation time
for $\tau=0.01$ h is quite long, i.e. $\tau_{\rm corr}\approx 2000$ h
for the $n=4$ case, even though the period fluctuations are substantial
as can be seen in Fig. \ref{dist} (See Appendix).

The scaling of $\tau_{\rm corr}$ given by Eq.(\ref{tau_corr}) can be
understood as the phase diffusion when the standard deviations of the
period distributions scales as $\sqrt\tau$ as shown in Fig. \ref{dist}.


\section*{Discussion}

We have examined the effects of molecular fluctuations in a biological
system on a simplified model of a circadian rhythm system, where there
are two types of fluctuation sources: (i) small numbers of molecules
involved and (ii) finite time scale of the gene regulation. The first
effect has been studied by
\cite{Gonze-2002-b}, assuming that the gene regulation time scale is
infinitesimal. In the present work, we focus on the second effect, 
\blue{i.e.,
in the case where the molecular numbers are large enough that the
fluctuation due to the first effect is negligible.}

\red{ We have developed a method to study this effect systematically by
introducing a new parameter $\tau$ to scale the gene regulation times.
We set $\tau$ to be the unbinding time of the transcription factor,
keeping the ratio of the binding to the unbinding rate constant.} We
performed numerical simulations for various values of $\tau$ without an
external entrainment of the 24-hour period.  As $\tau$ decreases, the
oscillation appears more deterministic; the width of the distributions
for the oscillation periods and the peak values scales with $\sqrt\tau$
and the correlation time for the correlations function scales with
$1/\tau$.  We have found \red{that the system is very sensitive to such
fluctuation, and demonstrated that the oscillation period fluctuates by
about 30 min even for very small $\tau=0.01$ h $\approx 30$ s in
comparison with its
period around 22 h.}
\blue{For the present parameter set, the nuclear protein concentration
$P_N/\Omega$ oscillates in the range $0\sim 5$ nM, therefore, the value
of $\tau\approx 30$ s for the unbinding time gives about 1 s for the
binding time.  These estimates may be tested with experimental data.}

\blue{ The 30 minutes period fluctuation is large for a circadian
system.  This sensitivity to the fluctuation in the gene regulation is
an interesting feature of the present simplified model.  Multiple
feedback loops with several phosphorylation steps found in actual
biological systems may be designed in such a way as to reduce this
sensitivity \citep{Ueda-2001,Kusakina-2012}.  This can be studied by
extending the present method.}

The correlation function for the protein oscillation fits to the damped
sinusoidal function very well, and the estimated correlation time
$\tau_{\rm corr}$ scales as $1/\tau$ in the small $\tau$ regime.  Such
decay in the correlation function is caused by the phase diffusion due
to fluctuations.  We estimate $\tau_{\rm corr}$ for the Hill coefficient
for the gene regulation $n=$4 and 1, and found that $\tau_{\rm corr}$'s
for $n=4$ are about 5 times larger than those for $n=1$; the fluctuation
effect is suppressed by the larger value of the Hill coefficient
\blue{by the cooperativity effects as in the case of
\cite{Gonze-2002-a,Gonze-2002-b}.}

\blue{ It is also interesting to find that the relative fluctuations for
the peak values are twice as large as those for the periods, namely, the
period is more stable than the amplitude.  This may be a reason why the
correlation time is quite long in spite of apparent fluctuations in the
oscillation.}

\red{ In the present work, we study only the case where the copy number
of the gene is 1.  However, there are typically a couple of genes in a
cell.  In the case of multiple genes in a cell, the fluctuation in each
gene cancels each other, therefore overall fluctuation will be reduced.
We confirmed by simulation that the fluctuation for a two-gene system
with $\tau$ is almost the same as that for a single gene system with
$\tau/2$. This is because the fluctuation cancellation by two genes
should be comparable with that by one gene that switches twice as fast.
Simulation data  are presented in Supplementary
material.}


\green{
The fluctuation indicated by our simulations may be compared with
previous experimental observations.}
%
%
Although circadian clocks are very accurate as a system, large
fluctuations have been observed in the oscillation of individual cells
of fibroblasts \citep{Nagoshi-2004} and cyanobacteria
\citep{Mihalcescu-2004} when they oscillate independently.  For both
cases, it is reported that the fluctuations are much larger for the
amplitude than those for the period.  In the latter case
\citep{Mihalcescu-2004}, the correlation time is estimated as long as
166$\pm$100 days in spite of apparent large fluctuations in the
amplitude.  Such a long correlation time corresponds to our case of the
gene regulation time scale $\tau=0.01$ h, which gives the $\tau_{\rm
corr}\approx 1950$ h.


Very little fluctuation is usually observed in circadian
systems; fluctuation in the period is typically less than 10
minutes \citep{Amdaoud-2007}, which is even smaller than the fluctuation
of 30 minutes that we obtained for the case $\tau=0.01$ h.  There are
some possible mechanisms to suppress molecular fluctuations.
(i) Cooperativity among cells: The present system models a single cell
behavior, but for the case of multicellular organisms, the cooperativity
among cells may exist and that should reduce the fluctuation in each
cell.  Actually, variability in each cell is much larger than that of
a whole system in the case of multicellular
organisms \citep{Liu-1997,Yamaguchi-2003,Herzog-2004,Carr-2005}.
(ii) Multiple feedback loops: Our model is a simplified core model for a
circadian system and consists of a single negative feedback
loop. However, it has been known that circadian systems typically
consist of multiple feedback
loops \citep{Zeng-1996,Leloup-1998,Glossop-1999,Blau-2001}, 
which could be
designed in the way to compensate the fluctuations in one loop by the
other.
(iii) Chemical oscillation without gene control: In the case of
cyanobacteria, it has been proposed that the circadian system consists
of proteins only and does not involve a gene
expression \citep{Tomita-2005}.  In such a system, the fluctuation
discussed in this work does not exist.

Other than circadian systems, there are some oscillations observed in
biology such as Hes1 oscillation during somite
segmentation \citep{Hirata-2002}, p53 oscillation after DNA damage by
gamma irradiation \citep{Geva-2006}, or oscillations in artificially
constructed systems \citep{Elowitz-2000,Atkinson-2003,Tigges-2009}.  In
these systems, the fluctuations are much more profound than circadian
systems, and part of the fluctuations should come from the gene
regulatory processes, for which the present analysis is applicable.

\blue{ In summary, we have developed a theoretical tool to study the
molecular fluctuation due to the finite transcription regulation time,
and have demonstrated that a symplified core model of circadian system
is sensitive to such fluctuation.  Our method can be extended to study a
more realistic system and can be utilized to clarify biological
significance of a detailed design of circadian system in terms of
stability against the molecular fluctuation.}


\appendix
    \renewcommand{\theequation}{A.\arabic{equation}}
    \setcounter{equation}{0} 
\section*{Appendix: Phase diffusion and correlation time}

In the appendix, we show that the correlation time $\tau_{\rm corr}$ in
the correlation function is proportional to $1/\tau$ when the period
distribution has the width proportional to $\sqrt\tau$.

Suppose the correlation function $C(t)$ is written as
\begin{equation}
C(t) = A \int_{-\infty}^\infty
\cos\bigl(\omega_0 t +\theta\bigr) P(\theta,t) d\theta
\label{C(t)}
\end{equation}
in terms of the average over the phase difference $\theta$ by the
distribution function $P(\theta,t)$ at time $t$.  Here, $\omega_0$ is
the average angular velocity given by $\omega_0=2\pi/T_0$ in terms of
the average period $T_0$.  Now, we assume that the phase distribution
can be approximated by the Gaussian distribution with
the standard deviation $\sigma_\theta(t)$, 
\begin{equation}
P(\theta, t) \approx
{1\over\sqrt{2\pi\sigma_\theta^2(t)}}
\exp\left[-{\theta^2\over 2\sigma_\theta^2(t)}\right],
\end{equation}
then, Eq.(\ref{C(t)}) may be estimated as
\begin{equation}
C(t)\approx A \cos(\omega_0 t) \,
\exp\left[-{1\over 2}\sigma_\theta^2(t)\right].
\label{C(t)-approx}
\end{equation}

Now, we estimate the phase distribution $P(\theta,t)$ as follows.  The
phase $\theta$ at the time $t=nT_0$ may be expressed as the sum of
$n$ phases accumulated by the time:
\begin{equation}
\theta = \sum_{i=1}^{n} 2\pi\left({1\over T_i}-{1\over T_0}\right)T_0
\approx -\sum_{i=1}^{n} 2\pi {\Delta T_i\over T_0},
\end{equation}
where $T_i$ is the $i$'th period (i.e. peak-to-peak interval) with
$T_i=T_0+\Delta T_i$, and we have assumed $\Delta T_i\ll T_0$ in the
last approximation.

If there is no correlation among $\Delta T_i$, then $\sigma_\theta(t)$
is given by
\begin{equation}
\sigma_\theta^2(t) \approx 
\left(2\pi{\sigma_T\over T_0}\right)^2 {t\over T_0},
\end{equation}
where $\sigma_T$ is the standard deviation of the period $T$.
We have replaced $n$ by $t/T_0$.

With Eq.(\ref{C(t)-approx}), this gives the form of
Eq.(\ref{damped-osci}) with $\theta_0=0$ and
\begin{equation}
\tau_{\rm corr} 
= \left({1\over 2\pi}\,{T_0\over \sigma_T}\right)^2 2T_0,
\end{equation}
thus if $\sigma_T\propto \sqrt\tau$, we obtain $\tau_{\rm corr}\propto
1/\tau$.

\vskip 1ex

\noindent
{\bf Acknowledgments: }
The authors would like to acknowledge Dr. Hiroshi Ito for informative
discussions.


\bibliographystyle{model2-names}
\bibliography{./references.bib}

\setlength{\topmargin}{-2.6cm}
\setlength{\textheight}{27cm}

\newpage


    \renewcommand{\theequation}{S.\arabic{equation}}
    \setcounter{equation}{0} 
    \renewcommand{\thefigure}{S.\arabic{figure}}
    \setcounter{figure}{0} 
    \renewcommand{\thetable}{S.\arabic{table}}
    \setcounter{table}{0} 
    \renewcommand{\thesection}{S.\arabic{section}}
    \setcounter{table}{0} 
\centerline{\parbox[t]{17cm}{
\begin{center}
\large\bf 
{\normalsize Supplementary material to} \\
``Transcription fluctuation effects 
on biochemical oscillations''\\
{\normalsize by Ryota Nishino, Takahiro Sakaue, and Hiizu Nakanishi}
\end{center} 
}}
\vskip 2ex

In this supplementary material, we present data to demonstrate the
$\sqrt\tau$-scaling of the distribution width and the multiple gene
effects more systematically.


\section{$\sqrt\tau$-scaling of the distribution width}

In the text, we show only three sets of data for the distributions of
the period and the peak value of $\rm [P_C]$ in order to demonstrate the
$\sqrt\tau$ scaling of the distribution width.  We examined this scaling
more systematically.

\begin{figure}[h]
\centerline{\includegraphics{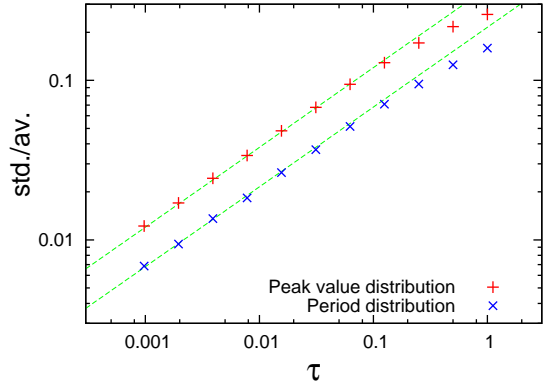}}
\caption{Scaling of the distribution width by $\tau$.  The ratio of the
standard deviation to the average is plotted against $\tau$ in the
logarithmic scale for the peak value of $\rm [P_C]$ and the period
distributions.  The dashed lines are the lines with the slope 0.5 fitted
to each data set.  }
\label{dist-scaling}
\end{figure}

In Fig.\ref{dist-scaling}, the ratios of the standard deviation to the
average are plotted against $\tau$ in the logarithmic scale.
The dashed lines denote the fitting lines with the slope 0.5, which
shows the $\sqrt\tau$ scaling of the distribution width of the
oscillation parameters in the small $\tau$ region.

This can be understood naturally; The on/off frequency of the gene is
proportional to $1/\tau$, thus the fluctuation in the total on/off-time
length scales with $\sqrt\tau$, from which we expect that the
distribution width of the oscillation parameters scales with $\sqrt\tau$
because the response in fluctuation should be linear in the small
input fluctuation limit.

\newpage
\section{Multiple gene effects}

In the paper, we only consider the case where the copy number of the
gene is one.  Here, we present the analysis for the case where the copy
number of the gene is $G_0$.  The transition rate for each reaction is
given by Table \ref{I}, where $G$ is the number of active genes.  Note
that the transition rate for the reaction 1 is scaled by $1/G_0$ in
order to give the same rate equation as before in the limit of $\tau\to
0$ and $\Omega\to\infty$.

In the $\tau\to 0$ limit, the time dependent average of $G$ denoted by
$G_{\rm Av}(t)$ is given by
\begin{equation}
G_{\rm Av}(t) = G_0{1\over 1 + ({\rm [P_N]}/K_I)^n},
\end{equation}
which gives the same rate equations for the concentrations if the
transition rate for the reaction 1 is scales by $1/G_0$.

\begin{figure}[h]
\centerline{\includegraphics{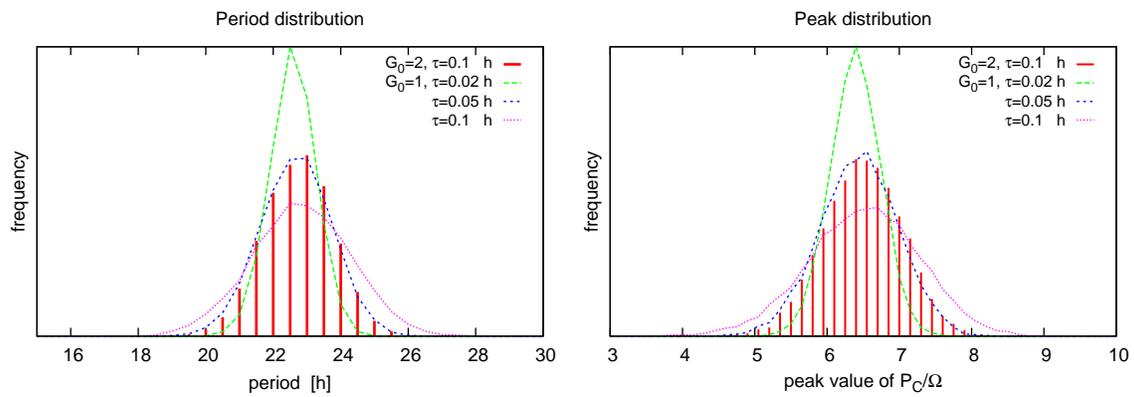}} 
\caption{ Period and $\rm
 [P_C]$ peak value distributions for $G_0=1$ and 2 with various values
 of $\tau$ in the $\Omega\to\infty$ limit.  
The other parameters are the same with those in the text:
$n=4$, $v_s=0.5$ nM h$^{-1}$, $K_I=2.0$ nM, $v_m=0.3$ nM h$^{-1}$,
$K_m=0.2$ nM, $k_s=2.0$ h$^{-1}$, $v_d=1.5$ nM h$^{-1}$,
$K_d=0.1$ nM, $k_1=0.2$ h$^{-1}$, $k_2=0.2$ h$^{-1}$.
}
\label{dist-2}
\end{figure}

To see the copy number effects, we performed Monte Carlo simulations for
$G_0=1$ and 2 with some values of $\tau$ in the $\Omega\to\infty$ limit.
Fig. \ref{dist-2} shows the distributions for the period and the peak
values of $\rm [P_C]$.  One can see the distribution for $G_0=2$ with
$\tau=0.1$ h agrees with that for $G_0=1$ with $\tau=0.05$ h quite
well for both of the distributions.

Actually, this can be understood in a simple way; For the copy number
$G_0=2$, the fluctuations in the activity of the two genes cancel each
other.  This cancellation should be comparable to the fluctuation
cancellation in the system with $G_0=1$ and the half time scale because
the gene activity switches between on and off twice as fast.

\begin{table}[h]
\centerline{\begin{tabular}{c@{\hspace{.5cm}}c c}
\hline
no. & reaction & transition rate
\\\hline\\
a & \parbox{3ex}{$G$\\ $P_N$}  $\longrightarrow$  
\parbox{7ex}{$G-1$\\$P_N -n$} &
$\displaystyle {1\over\tau}\,\left({P_N\over K_I\Omega}\right)^n G$ 
\\\\
b & \parbox{3ex}{$G$\\ $P_N$}  $\longrightarrow$  
\parbox{7ex}{$G+1$\\$P_N +n$} & 
$\displaystyle {1\over\tau}\,(G_0-G)$ 
\\\\\hline\\
1 & $M$  $\longrightarrow$  $M+1$ &
$\displaystyle v_s\Omega\, {G\over G_0}$ 
\\\\
2 & $M$  $\longrightarrow$  $M-1$ & 
$\displaystyle v_m\Omega\,{M/\Omega\over K_m+M/\Omega}$ 
\\\\
3 & $P_C$  $\longrightarrow$  $P_C+1$ & $k_s M$ 
\\\\
4 & $P_C$  $\longrightarrow$  $P_C-1$ &
$\displaystyle v_d\Omega\,{P_C/\Omega\over K_d+P_C/\Omega}$ 
\\\\
5 & \parbox{0.5cm}{$P_C$\\$P_N$}  $\longrightarrow$  
\parbox{1.1cm}{$P_C-1$\\$P_N+1$} &  $k_1 P_C$ 
\\\\
6 & \parbox{0.5cm}{$P_C$\\$P_N$}  $\longrightarrow$  
\parbox{1.1cm}{$P_C+1$\\$P_N-1$} & $k_2 P_N$ 
\\\\\hline
\end{tabular}}
\caption
{Reaction table for a simplified circadian system in the case where the
 copy number of the gene is $G_0$.}\vskip 3ex
\label{I}
\end{table}

\end{document}